\documentclass[12pt]{article}
\usepackage{graphicx,amsmath}
\usepackage{lineno,color}
\usepackage{units}

\newlength{\dinwidth}
\newlength{\dinmargin}
\renewcommand{\vec}[1]{\boldsymbol{#1}}

\def\lapproxeq{\lower .7ex\hbox{$\;\stackrel{\textstyle                                                    
<}{\sim}\;$}}                                                    
\def\gapproxeq{\lower .7ex\hbox{$\;\stackrel{\textstyle                                                    
>}{\sim}\;$}}                                                    
\def\be{\begin{equation}}                                                    
\def\ee{\end{equation}}                                                    
\def\bea{\begin{eqnarray}}                                                    
\def\eea{\end{eqnarray}}
\def\b{\vec{b}}
     
\def\q{\vec{q}} 
\def\p{I\!\! P}

\def\sh{\hat s}
\def\sh2{{\hat s}^2}

\begin{document}
\titlepage                                                    
\begin{flushright}                                                    
IPPP/24/24  \\                                                    
\today \\                                                    
\end{flushright} 
\vspace*{0.5cm}
\begin{center}                                                    
{\Large \bf Oscillations in the elastic high energy amplitude}\\
\vspace*{1cm}
                                                   
 P. Grafström$^a$,
A.D. Martin$^b$ and M.G. Ryskin$^{c}$ \\                                                    
                                                   
\vspace*{0.5cm}                                                    
$^a$ Università di Bologna, Dipartimento di Fisica, 40126 Bologna, Italy\\
$^b$ Institute for Particle Physics Phenomenology, University of Durham, Durham, DH1 3LE \\                                                   
$^c$ Petersburg Nuclear Physics Institute, NRC Kurchatov Institute, Gatchina, St.~Petersburg, 188300, Russia

\vspace*{1cm}

\begin{abstract}

  We discuss the oscillations in the elastic $pp$ differential cross section seen in the TOTEM data at $\sqrt s=13$ TeV on the top of the usual smooth behaviour.
\end{abstract}

\vspace*{1cm}  
  \vspace{1cm}

 \vfill

 
\end{center}
 \newpage

\section{Introduction}

The possibility of oscillations in the high energy elastic scattering was predicted theoretically long ago by Anselm and Dyatlov in \cite{AD}. These oscillations are caused by the alternative signs of the multi-Pomeron contributions. At this time the theory of strong interactions was based on three tenets: unitarity, crossing and analyticity, which arise from causality.

In fact, the hope had been that the implementation of these three tenets would reveal the hadron spectrum (see, for example, \cite{Ch}). It is relevant to note that the behaviour of the particles in the complex angular momentum plane is given by their Regge pole trajectory, see, for example, \cite{C}. Unfortunately this boot-strap idea to generate the hadron spectrum was torpedoed by the revelation of the CDD ambiguity \cite{CDD}.

As mentioned above, the dominance of the vacuum (Pomeron) singularity and the secondary  Regge poles means that the oscillations should occur in elastic $pp$ scattering. They are features of unitarity, crossing and analyticity. Indeed, these oscillations may be visible in very high energy elastic $pp$ (and proton-nuclear) scattering data in the very forward direction. Here we will quantify the  oscillation signals that are expected in the experiments which are underway at the LHC at CERN.\\

Later using axiomatic field theory (plus some additional quite reasonable constraints) it was shown~\cite{AKM}  that the scattering amplitude must have infinitely many zeros in the forward direction. In spite of the fact that these zeros are placed in the complex plane (but close to the real $t$ axis) such properties of the scattering amplitude may lead to oscillations of the differential cross section in the forward  
direction.~\footnote{It is not excluded  ``that the zeros might not be close enough to
the physical region to produce oscillations"~\cite{AM}. However  the first diffractive dip should be seen.}\\

The expected oscillations were extracted from the very precise TOTEM 13 TeV data~\cite{T13} by Selyugin in \cite{Sel} (see e.g. Fig.5) where the HEGS~\cite{HEGS,SEL1} model was used to describe the smooth behaviour of the amplitude.
 Similar oscillations were observed in~\cite{Per} (see Figs. 2b,3a) using a Phillips-Barger Regge parameterization ~\cite{PB} plus an oscillating term. That is the observed oscillations at 13 TeV do not depend on a particular parameterization used to describe the smooth amplitude behaviour. At  smaller energies the accuracy of the data are not sufficient to make a  definite conclusion - whether there are some oscillations or not~\cite{Per}. Note that the amplitude of observed oscillations is small - only about 0.5\% of the corresponding cross section.\\
 
 At first sight, these observations could be regarded as an exciting confirmation of theory. However, as we will quantify below, the problem is that the oscillations were observed at  low $|t|$ values, before the first diffractive dip.\\

 In Section 2 we recall the origin of oscillations produced by the multi-Pomeron diagrams generated by the two-particle $s$-channel unitarity equation. In Section 3 we consider the amplitude in the impact parameter, $b$, representation. At high energies, $\sqrt s$, this is equivalent to the partial wave, $l$, expansion ($l=b\sqrt s/2$). In conclusion, we emphasize that the oscillations seen in \cite{Sel,Per} should be caused by some irregularity of the amplitude at a very large $b\sim 6$ fm values. This looks quite strange.

 \section{Oscillations due to multi-Pomeron exchange}
\begin{figure} [t]
\begin{center}
\includegraphics[trim=0.0cm 0cm 0cm 0cm,scale=0.43]{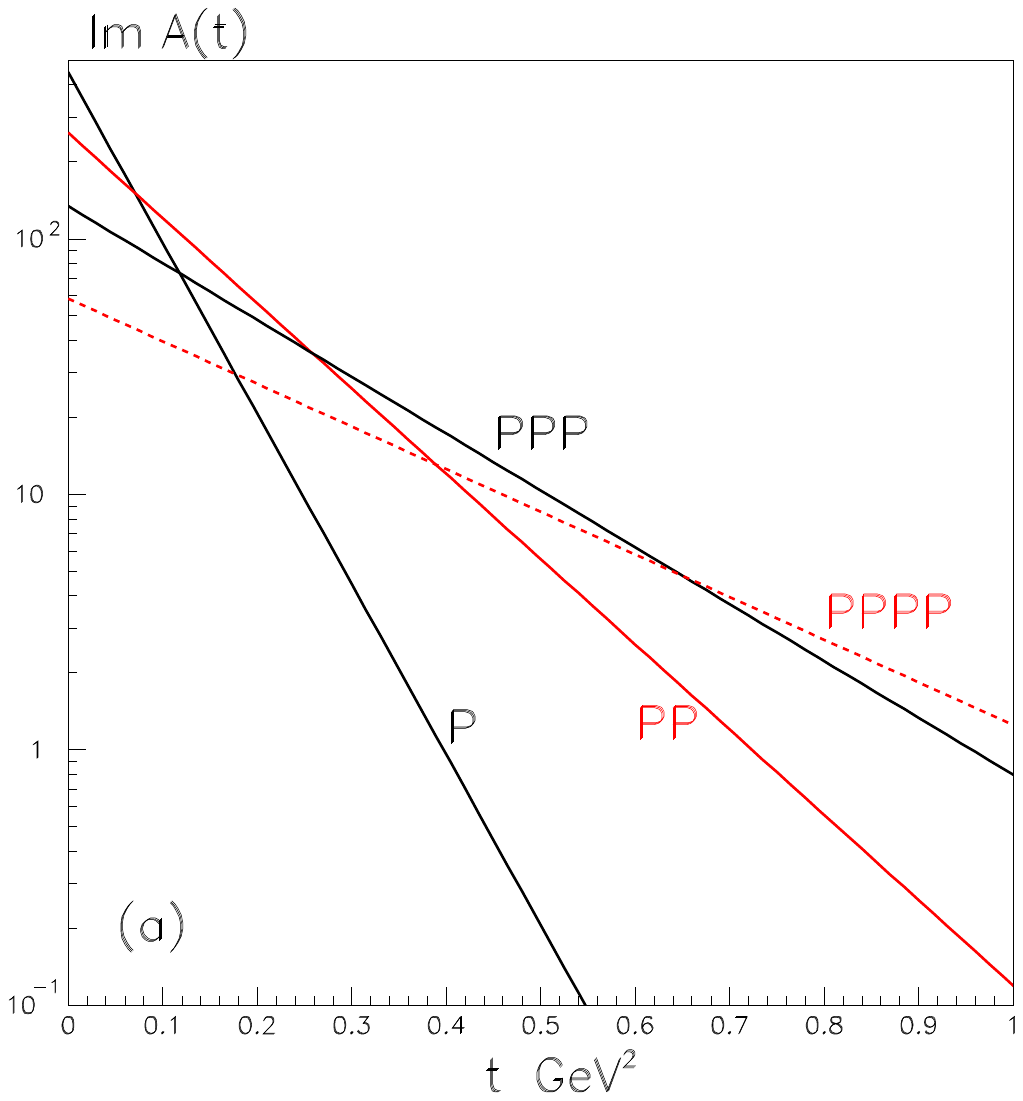}
\hspace{-1.5cm}
\includegraphics[trim=0.0cm 0cm 0cm 0cm,scale=0.43]{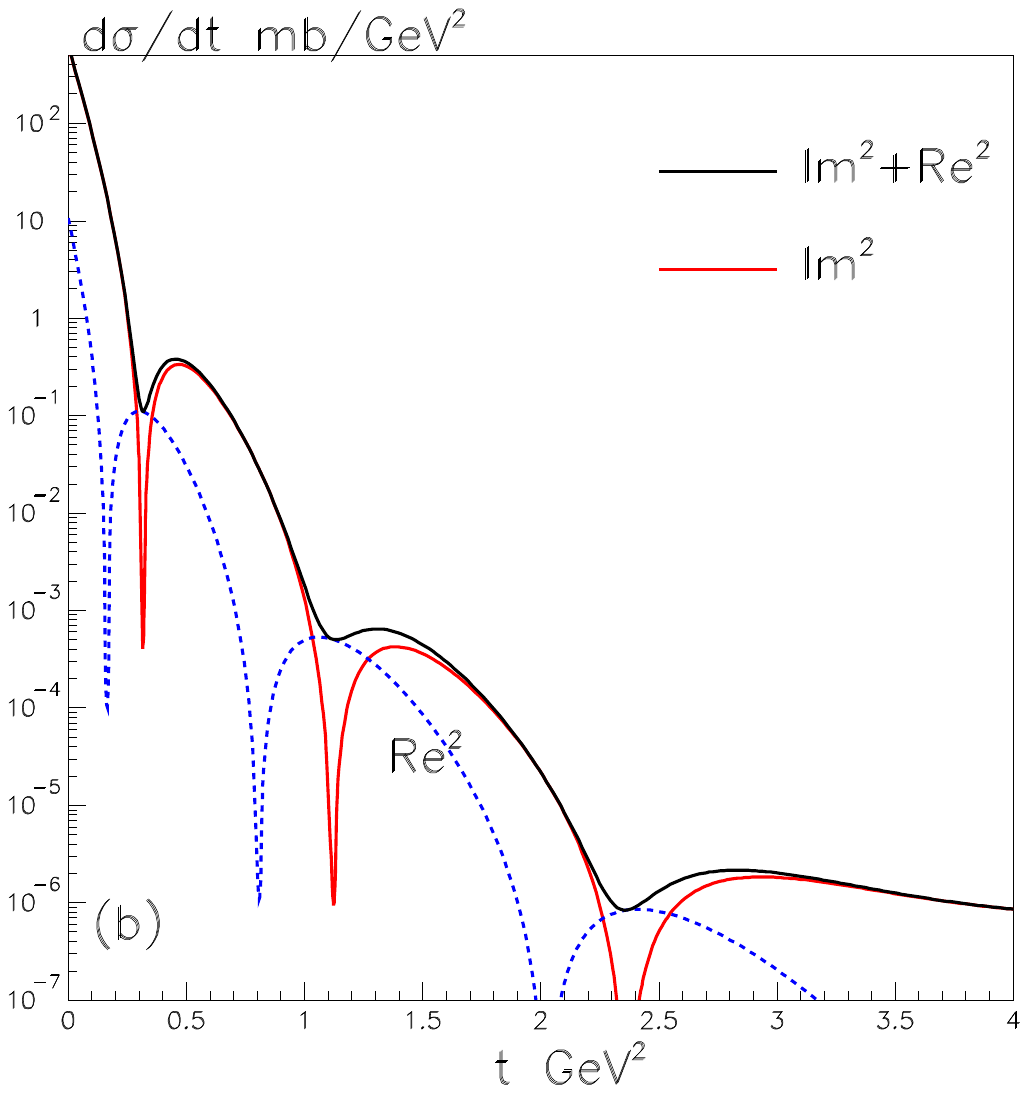}
\caption{\sf (a) The contribution of the individual multi-Pomeron diagrams to the elastic cross section; negative contributions are shown in red.  (b) The elastic cross section generated by the sum of multi-Pomeron diagrams. The contribution of the imaginary part of amplitude is shown in red, while the contribution of the real part is show by the blue dashed curve.
}
\label{1}
\end{center}
\end{figure}

Starting with  one Pomeron, $\p$ , exchange  we discover that in order to satisfy unitarity
\be
2\mbox{Im}A_{el}(q)=\frac 1{8\pi^2 s}\int d^2k A^*_{el}(k)A_{el}(q-k) + \sum_{n\neq i} A^*_{in}A_{ni}
\label{u1} 
\ee
one has to add the two, three and other multi-Pomeron diagrams~\cite{RFT,C} (here $\sum_{n\neq i}A^*_{in}A_{ni}=G_{inel}$ corresponds to the contribution of inelastic states).  In the Pomeron exchange amplitude the  imaginary part dominates. Thus, in the first approximation we may neglect the real part and the elastic amplitude becomes the sum of terms with alternative sign
\be
A_{el}=\p-\p\p+\p\p\p-\p\p\p\p+...
\ee
In the case  of a few Pomerons the total momentum transferred, $t=-q^2_t$, can be divided between the Pomerons. Thus the $q_t$ dependence of the diagrams  becomes flatter. Indeed, if the one-Pomeron exchange amplitude is $A_1(t)\propto \exp(B_1t)$,  then for the two Pomeron contribution we have $A_2(t)\propto \exp(B_1t/2)$, while for $A_3(t)\propto \exp(B_1t/3)$ and so on.  This is shown in Fig.\ref{1}a. 
The resulting cross section is presented in Fig.\ref{1}b by the black curve. The red curve shows the contribution of the imaginary part of the final amplitude and the dips indicate the points where
 $\mbox{Im}A_{el}(t)=0$. These dips are filled by the real part which has its own zeros at a slightly different $t$ values.\\
  
 Note that the Figure has a pure illustrative character. We have assumed a pure exponential behaviour of the Pomeron exchange amplitudes as it was done in the original oscillation paper~\cite{AD}. The real $t$ dependence is more complicated. However here we have chosen  parameters which reasonably reproduce the behaviour of elastic $pp$ aamplitude observed by TOTEM at 13 TeV. In particular the total cross section $\sigma_{tot}=110$ mb and the elastic slope $B_{el}=20.4$ GeV$^{-2}$. More details of the parametrization used are described in Appendix.\\

 Going to  complex $t$ it becomes possible to nullify the entire amplitude, that is the imaginary and real parts vanish simultaneously. These are the zeros discovered in the AKM paper~\cite{AKM}. Clearly the presence of these zeros (mainly the vanishing of $\mbox{Im}A_{el}(t$)) reveal themselves as oscillations in the $t$ dependence of the differential cross section.\\

A similar behaviour of differential cross section was observed in proton-nuclei collisions. It is well described by Glauber theory ~\cite{Gl} which accounts for the possibility of proton rescattering on the different nucleons in nucleus (see e.g. Fig.s 9,19,24 of \cite{ABV}).

\section{Impact parameter representation}
A convenient way to account for $s$-channel unitarity is to work in the impact parameter, $b$, representation. At high energies the value of $b$ is fixed with  good ($\sim 1/s$) accuracy and the unitarity equation reads
\be 
\label{ub}
2\mbox{Im}A_{el}(b)=|A_{el}(b)|^2+G_{inel}(b)\ .
\ee
The solution of eq.(\ref{ub}) takes the form
\be 
\label{eik}
 A_{el}(b)=i(1-e^{-\Omega(b)/2}),\quad\quad G_{inel}(b)=1-e^{-\Omega(b)}\ ,
\ee
where the factor $e^{-\Omega(b)}$ is the probability to have no inelastic interactions.\\

The so-called opacity, $\Omega(b)$, describes one Pomeron exchange. It can be calculated as
\be 
\label{op}
\Omega(b)=\frac{-i}{4\pi^2s}\int d^2q_tA_1(q_t)e^{i\b\q}~=~\frac{g^2_N}{4\pi B_1}\left(\frac s{s_0}\right)^{\alpha_P(0)-1}e^{-b^2/4B_1}
\ee
for a pure exponential one-Pomeron amplitude 
\be 
\label{6}
A_1~=~isg^2_N\left(\frac s{s_0}\right)^{\alpha_P(0)-1}e^{B_1t}\ ,
\ee
as used in Section 2.

Here $g_N$ is the Pomeron-proton coupling and $\alpha_P(0)$ is the Pomeron intercept.\footnote{For simplicity we consider just the imaginary amplitude $A_1$. In a general case the opacity $\Omega$ should be complex to account for the real part of $A_1$.}

The full amplitude in $t$ representation is given by the inverse transform of (\ref{eik})
\be 
\label{amp}
A_{el}(t=-q^2_t)=is\int A_{el}(b)J_0(bq_t)d^2b\ ,
\ee
where $J_0(x)$ is the Bessel function.\\

\begin{figure} [t]
\begin{center}
\vspace{-5cm}
\includegraphics[trim=0.0cm 0cm 0cm 0cm,scale=0.8]{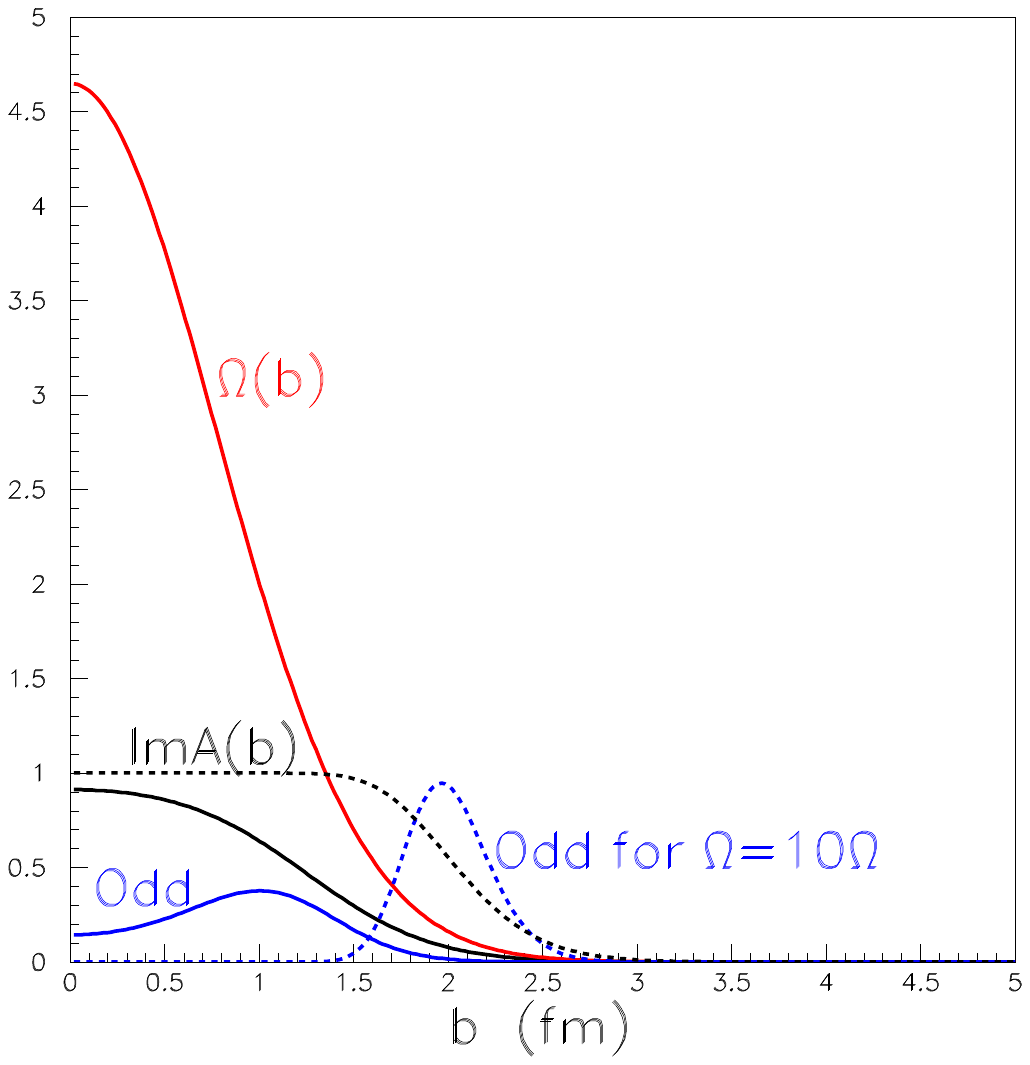}
\caption{\sf Impact parameter $b$ dependence of opacity (upper red curve) and the elastic amplitude (continuous black curve). The amplitude in the case of 10 times larger opacity is shown by black dashed line. The number 10 is arbitrary; it was chosen so as to make the effect  sufficiently visible. As an example by blue curves are shown the additional contribution called ``Odd" which originally has the pure exponential $b$ dependence but was distorted by the Pomeron screening. Blue dashed curve corresponds to 10 times larger opacity.
}
\label{2}
\end{center}
\end{figure}

The typical $b$ dependence of opacity is shown in Fig.\ref{2} by the red curve and the corresponding elastic amplitude by the black continuous curve.
For a much larger opacity the elastic amplitude starts to look like the black disk shown by the dashed black curve. In this case the expected $t$ behaviour of differential cross section $d\sigma/dt$ becomes similar to the usual diffraction on a black disk.

Note however that in order to enlarge the opacity, for example by the factor 10 in comparison with that at 13 TeV, when indeed the amplitude starts to look as the black disk (see the black dashed curve in Fig.\ref{2}) we have to go to extremely large energy. Assuming the Pomeron 
intercept $\alpha_P(0)=1+0.12$ we need an unrealistically high energy $\sqrt s\sim 2\cdot 10^5$ TeV.


At the present LHC energies the $b$ dependence of the  elastic amplitude is sufficiently smooth and the inverse transform (\ref{amp}) generates only one diffractive dip
(at $t\simeq -0.46$ GeV$^2$ for 13 TeV); as shown in the detailed analysis leading to Fig.2 of \cite{KMR}\\

Now let us add some small additional interaction 
$\delta\Omega(b)$. In the elastic process this additional interaction will be screened by the Pomerons leading to
\be 
\delta A_{el}(b)=\frac{\delta\Omega(b)}{2}e^{-\Omega(b)/2}\ .
\ee
The corresponding contribution 
 is shown in Fig.\ref{2} by the blue curves which are marked by ``Odd"; (the case of 10 times larger
  $\Omega$ is shown by a dashed line).
For the $b$ dependence of $\delta\Omega(b)\propto exp(-b^2/4B')$ we use the same exponential form but with   slope $B'$ two times smaller than that for elastic amplitude. The normalization of $\delta\Omega$ is arbitrary, just to make it clearly seen in Fig.\ref{2}.

 In the extreme case  (close to a black disk) of very large $\Omega$, this additional contribution survives only at the edge of the disk (somewhere near $b\simeq R$). 
After the transform (\ref{amp}) we then would expect the oscillations caused by a Bessel function $J_0(Rq_t)$.
If the oscillations observed in TOTEM 13 TeV data are of {\em this} origin then they should be produced by some irregularity of the $A_{el}(b)$ amplitude at rather large distances.

\section{Discussion}
\begin{figure} [t]
\begin{center}
\vspace{-5cm}
\hspace{-1.25cm}
\includegraphics[trim=0.0cm 0cm 0cm 0cm,scale=0.5]{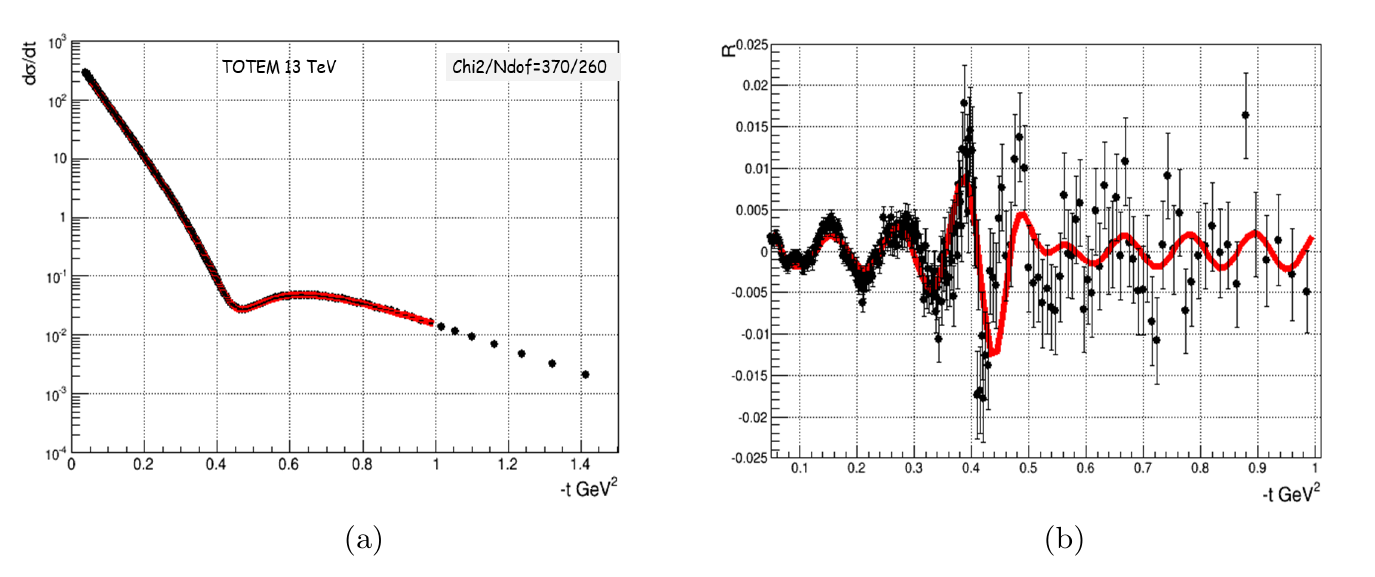}
\caption{\sf (a) TOTEM data at 13 TeV \cite{T13} as a function of $-t$. The data have been fitted with
the Phillips-Barger model \cite{PB} and adding an oscillatory term. 
(b) The ratio $R=d\sigma_{el}/dt(\mbox{data})/d\sigma_{el}/dt(\mbox{P-B fit})-1$ 
as a function of $-t$. The red curve represents
the oscillatory contribution found in the fit.}
\label{3}
\end{center}
\end{figure}

Theoretically the multi-Pomeron oscillations must exist \cite{AD,AKM}. At least the first diffractive dip is well visible in $pp$ scattering  (see \cite{KMR} and the footnote 1). 
To set the scene, in Fig.\ref{3} we reproduce Fig.2 of \cite{Per} which shows a description of the 13 TeV TOTEM data.
 Note that the oscillations occur down to low $|t|$, below the first diffractive dip. 
 Clearly these are not the oscillations generated by the multi-Pomeron diagrams which are discussed in \cite{AD,AKM}. Indeed we observe at least three oscillations before the dip at $-t\simeq 0.46$ GeV$^2$
 while the predicted AD-AKM~\cite{AD,AKM} oscillations  generated by multi-Pomeron diagrams start {\em after} the dip, as  seen in Fig.\ref{1}.
 
Recall also, that in the proton-nuclei scattering the density variation placed at the distance of about the ion/nucleus radius lead to some additional contribution (oscillations) only at the angles ($t$ values) larger than the first diffractive dip; see e.g. Fig.s 19,20 of \cite{ABV}. \\

Returning to elastic $pp$ interactions, we see from Fig.\ref{3}(b), that up to the first dip at $|t|=0.46$ GeV$^2$ there are three periods of oscillations. That is the Bessel function argument $bq_t$ should be about $6\pi$. In other words the characteristic radius 
 $$b = R = \frac{6\pi}{\sqrt{0.4\mbox{GeV}^2}} = 5.9\mbox{ fm}.$$

This is an unusually large value. It corresponds, in terms of the Yukawa coupling, to a mass of about $m$ = 30 MeV. We tried to add to the elastic amplitude an additional vector-particle exchange with mass $m$ = 10 - 50 MeV. No such oscillations were observed. Also no similar oscillations were obtained when we add to the proton coupling a large radius cloud, that is an extra form-factor like $
1/(t - m^2 )$ or $1/(t - m^2 )^ 2$ . 
 The problem is that on the one hand we need a small mass to go to large impact  $b \sim 6$ fm, while on the other hand (with the present quite smooth $A_{el}(b)$ behaviour of the main amplitude) we need a large mass to produce the sharp form of this new contribution.
 
\begin{figure} [t]
\begin{center}
\vspace{-5cm}
\hspace{-2. cm}
\includegraphics[trim=0.0cm 0cm 0cm 0cm,scale=0.7]{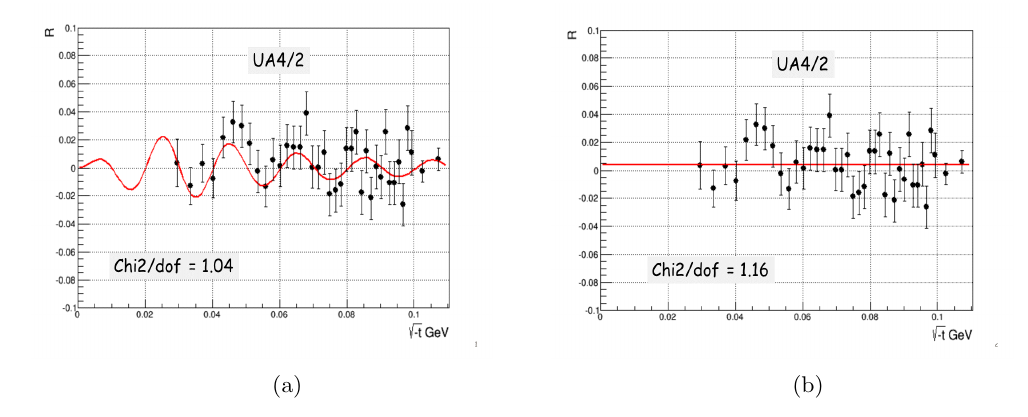}
\caption{\sf (a) UA4/2 data at 541 GeV \cite{UA4} as a function of $\sqrt{-t}$.  The data have been fitted by the exponent ($A(t)\propto e^{Bt/2}$) adding (a) or without (b) the oscillatory term. The ratio $R=d\sigma_{el}/dt(\mbox{data})/d\sigma_{el}/dt(\mbox{exp - fit})-1$. }
\label{4}
\end{center}
\end{figure}

 To be complete we have to add that  similar low-$t$ oscillations were observed (but with a lower statistical significance) long ago in the UA4/2 experiment~\cite{UA4} at $\sqrt s=541$ GeV. The discussion can be found in \cite{GNS,PG}. In this (UA4/2) case the first three oscillations are ``observed" (with, as noted,  low statistical  significance) at $-t<0.01$ GeV$^2$ -- 40 times smaller than that at 13 TeV (see Fig.\ref{4} taken from \cite{PG}). That is the corresponding irregularity of the amplitude should be placed at $b\sim 35$ fm ! Note that the AKM zeros \cite{AKM} come from the same $s$ channel two-particle unitarity
equation that generates the multi-Pomeron diagrams. Therefore they cannot produce oscillations for $|t| < 0.01$ GeV$^2$, contrary to the approach given in \cite{GNS}.\\

Recall that the signal for oscillations is weak, so it is very difficult to identify it by experiment. On the other hand if oscillations like those in the very high statistics sample of TOTEM \cite{T13} are found in the corresponding ATLAS/ALFA data with equally high statistics , then it means
 that we face {\em a completely new and very interesting phenomena}~\footnote{In the recent paper~\cite{PT} the TOTEM data  were analysed using as a seed cross section the model~\cite{MN}. The authors  did not observe a clear oscillation signal and state that searching for such a subtle effect looks premature. On the other hand the parameterization \cite{MN} contains too many free parameters. So it is not excluded that  part of oscillations were absorbed  by the seed amplitude.}.

\section*{Appendix: Simplified  Pomeron exchange amplitude}
To calculate the curves in Fig.s \ref{1},\ref{2} we parametrize the one-Pomeron exchange amplitude as
\be 
A_1(t)~=~sg^2_N(t)\eta_P(t)\left(\frac s{s_0}\right)^{\alpha_P(t)-1}\ ,
\label{9}
\ee
where the Pomeron signature factor $\eta_P(t)=-exp(-i\pi\alpha_P(t)/2)$ and the Pomeron trajectory $\alpha_P(t)=1+\Delta+\alpha'_Pt$. The Pomeron-nucleon coupling $g_N(t)=g_0e^{B_0t/2}$.

Correspondingly for the opacity $\Omega(b)$ we get\footnote{Actually in Fig.\ref{2} we plot the real part of $\Omega$.}
\be
\label{10}
\Omega(b)=\frac{-i}{4\pi^2s}\int d^2q_tA_1(q_t)e^{i\b\q}~=~-i\frac{g^2_N(0)}{4\pi B_\Omega}\eta_P(0)\left(\frac s{s_0}\right)^{\alpha_P(0)-1}e^{-b^2/4B_\Omega}\ ,
\ee 
where $B_\Omega=B_0+\alpha'_P(\ln(s/s_0)-i\pi/2)$.\\
Equations (\ref{9},\ref{10}) become identical to (\ref{6},\ref{op}) if we replace $B_1$ by $B_\Omega$ and the factor $i$ in (\ref{6}) by the signature  $\eta_P(0)$.

The parameters of Pomeron trajectory ($\Delta=0.13$ and $\alpha'_p=0.052$ GeV$^{-2}$) were taken from the fit to collider data presented in~\cite{KMR} while the values of coupling $g^2_N(0)=15.2$ mb and slope $B_0=13.4$ GeV$^{-2}$
are chosen to reasonably reproduce the TOTEM 13 TeV $\sigma_{tot}=110$ mb and elastic slope $B_{el}=20.4$ GeV$^{-2}$.

  

\end{document}